\documentclass[preprint2]{aastex701}

\usepackage{graphicx} 

\usepackage{orcidlink}
\usepackage{amsmath}

\shorttitle{Nuclear Radio Emission in NGC~7479}
\shortauthors{Laine et al.}

\makeatletter
\renewcommand{\frontmatter@title@above}{}
\makeatother

\accepted{April 17, 2026}
\submitjournal{AJ}

\begin{document}

\title{Very Long Baseline Interferometry Search for Nuclear Radio Continuum Emission in the Barred Spiral Galaxy NGC~7479}

\author[0000-0003-1250-8314]{Seppo Laine} 
\affiliation{IPAC, Mail Code 314-6, Caltech, 1200 E. California Blvd., Pasadena, CA 91125, USA}

\email{seppo@ipac.caltech.edu}

\author[0000-0003-3168-5922]{Emmanuel Momjian}
\affiliation{National Radio Astronomy Observatory, P.V.D. Science Operations Center, P.O. Box O, 1011 Lopezville Road, Socorro, NM 87801-0387, USA}

\email{emomjian@nrao.edu}

\author[0000-0001-9194-7168]{Emilia J\"{a}rvel\"{a}}
\affiliation{Department of Physics and Astronomy, Texas Tech University, Box 41051, Lubbock, 79409-1051, TX, USA}

\email{ejarvela@ttu.edu}

\author[0000-0002-4892-9586]{Thomas P. Krichbaum}
\affiliation{Max-Planck-Institut f\"{u}r Radioastronomie, Auf dem H\"{u}gel 69, 53121, Bonn, Germany}

\email{tkrichbaum@mpifr-bonn.mpg.edu}

\author[0000-0002-9214-4428]{S. Komossa}
\affiliation{Max-Planck-Institut f\"{u}r Radioastronomie, Auf dem H\"{u}gel 69, 53121, Bonn, Germany}

\email{skomossa@mpifr-bonn.mpg.edu}

\author[0000-0002-3365-8875]{Travis C. Fischer}
\affiliation{AURA for ESA, Space Telescope Science Institute, 3700 San Martin Drive, Baltimore, MD 21218, USA}

\email{tfischer@stsci.edu}

\author[0000-0001-7658-1065]{Thomas G. Pannuti}
\affiliation{Department of Engineering Science, 123 Lappin Hall, Morehead State University, 150 University Drive, Morehead, KY 40351, USA}

\email{ t.pannuti@moreheadstate.edu}

\begin{abstract}

We have obtained very high angular resolution (a few milliarcseconds or sub-parsec scale) Very Long Baseline Array (VLBA) and European Very Long Baseline Interferometry (VLBI) Network (EVN) radio continuum images of the nucleus in the barred spiral galaxy NGC~7479, to search for possible nuclear emission on parsec scales. The observations were taken using phase referencing. Previous Karl G. Jansky Very Large Array (VLA) and Multi-Element Radio Linked Interferometer Network (MERLIN) observations revealed a large jet-like structure, apparently emanating from the nucleus, and unresolved nuclear emission at 0.1 arcsecond ($\approx$~15 pc at the assumed distance of 32 Mpc) scale, respectively. Our sensitive new VLBA and EVN images resolve the previously unresolved nuclear source and reveal two distinct emission regions (VLBI components) that are separated by about 30 milliarcseconds. We also report an apparent change in separation of the two main emission regions over the ten years between EVN and VLBA observations, implying relativistic radio jet motion or changes in shock illumination of gas by a nuclear wind. We measure the spectral indices and brightness temperatures of the VLBI components, and discuss possible physical causes of the observed emission.

\end{abstract}

\section{Introduction} {\label{intro}}

\setcounter{footnote}{0}

A lot of progress has been made in understanding the triggering and driving mechanisms of powerful extragalactic jets (by jets we mean collimated, relativistic flows that generate the radio emission directly) by recent Very Long Baseline Interferometry (VLBI) RadioAstron and Event Horizon Telescope observations \citep[e.g.,][]{vega2020,roder2025}. To date, most of the efforts to understand the origin of extragalactic jets have concentrated on large-scale (thousands of parsecs) jets in distant, radio-loud quasars and blazars that are thought to be associated with single \citep[e.g.,][]{meier2001,mckinney2006} or binary \citep[e.g.,][]{merritt2002,komossa2006,valtonen2007,britzen2018} supermassive black holes (SMBHs).

An alternative to the jet explanation for large-scale jet-like structures in galaxies with radio-quiet active galactic nuclei (AGNs) is a wide-angle, non-relativistic bi-polar outflow, emerging from an AGN (a ``wind''). These winds can trigger synchrotron emission by mechanical shocks when they impact the medium surrounding the nucleus from sub-parsec to kiloparsec scales \citep[e.g.,][]{smith2020,fischer2021,fischer2023,chen2024,harrison2024,fischer2025}. Radio continuum emission from radio-quiet AGNs has been reviewed by \citet{panessa2019}.

The study of the physics of how the kpc-scale jets or outflows are triggered and driven is hampered by current observational spatial-resolution limitations. A few nearby elliptical galaxies with kiloparsec (kpc)-scale jet-like features that offer higher physical spatial resolution exist, most notably M87 \citep[e.g.,][]{owen1980,lu2023} and Cen A \citep[e.g.,][]{israel1998,janssen2021}. 

Among nearby (within 100 Mpc) disk galaxies large-scale radio jet-like features are rare. The best known example is NGC~4258 (distance $\lesssim$~8 Mpc), for which a minor merger has been postulated as the origin of the several kpc scale trailing jet-like arm structure \citep{taniguchi1996}. In NGC~4258 the emission from anomalous radio arms is understood to be the result of the interaction between the disk of the galaxy and radio jets \citep[e.g.,][]{krause2007} or of the interaction of an AGN outflow with the interstellar medium near the disk plane \citep{fischer2025}. Evidence for the minor merger as a possible triggering mechanism is lacking, although there may be tidal interaction, seen in \ion{H}{1} gas \citep{zhu2021}.

\begin{figure*}[t]
  \begin{center}
    \includegraphics[width=0.8\linewidth]{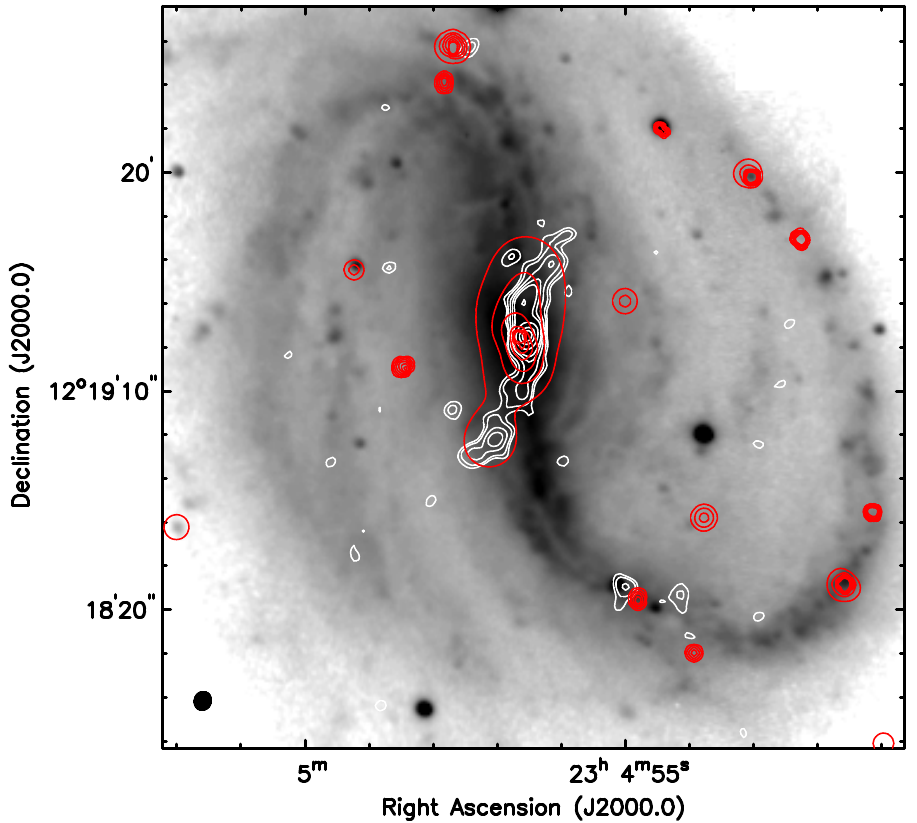}
  \end{center}
  \caption{\label{radiocontlarge}\small $V$-band grayscale image with white contours of VLA 21 cm emission and red contours of Chandra soft X-ray band emission of NGC~7479. The VLA 21 cm B-configuration radio continuum emission is shown with contours at 1.5$\times$10$^{-4}$ $\times$ (3, 4, 6, 8, 10, 16, 24, 30) Jy~beam$^{-1}$. The VLA beam size (4\farcs 30$\times$3\farcs 85, position angle $-19\degr$) is shown with the black ellipse at the lower left corner. The soft X-ray emission from archival Chandra X-ray Observatory data is shown with contours in arbitrary units. The $V$-band image was taken with the Isaac Newton Telescope's Prime Focus Cone Unit using 600 second exposures on 1994 September 22. An unrelated VLA background point source about 30 arcseconds to the east of the nucleus was removed in the $u-v$ space before imaging.}
\end{figure*}

NGC~7479 is the next nearest
($z$ = 0.007925, distance 32~Mpc\footnote{We use the same distance as adopted by \citeauthor{laine1998} (\citeyear{laine1998}). At this distance, 1 milliarcsecond (mas) corresponds to 0.15~pc.}; $H_{0}$ = 73~km~s$^{-1}$ and $\Omega_{\rm M}$ = 0.3)
spiral galaxy with almost bisymmetric, several kpc scale, slightly out of the plane, radio jet-like structures \citep{laine1998,laine1999,laine2008}. These structures are also seen in X-rays \citep[e.g.,][]{fadda2021}, surrounded by young star formation regions in a shock front morphology.
A rough upper estimate on the star formation rate in the bar region due to the jet-like feature is obtained by using the integrated 21-cm flux density from \citet{laine1998} as a substitute for the 20-cm integrated flux density and the relationship from \citeauthor{bell2003} (\citeyear{bell2003}; see also \citeauthor{murphy2011} \citeyear{murphy2011}). Using this integrated flux density and this relationship, we compute a star formation rate of 2.7 M$_{\sun}$~yr$^{-1}$ in the bar region. On the sky these ``jets'' appear as "S"-shaped arms, contrasting to the "Z"-shaped stellar spiral arms (as seen in Figure~\ref{radiocontlarge}). 

Overall, the large-scale radio continuum structure in NGC~7479 is reminiscent of the one seen in the luminous infrared galaxy VV~340a that has been modeled by a precessing weak radio jet \citep{kader2026}. The strong radio continuum emission in NGC~7479 has allowed the study of the disk magnetic field in detail \citep{laine2008}. These authors showed that the magnetic field B vectors are aligned along the S-shaped, kpc-scale radio continuum jet-like structure. The earlier, roughly one-arcsecond resolution Karl G. Jansky Very Large Array (VLA) observations at 6 and 20~cm \citep{ho2001,laine2006}, revealed an unresolved nuclear point source, in addition to radio continuum emission to the north, starting at a few arcsecond or a few hundred parsec distance from the nucleus. 

NGC~7479 is also a good candidate for a recent (minor) merger \citep[e.g.,][]{quillen1995,laine1998,laine1999,martin2000}. Thus, this galaxy provides us with one of the best laboratories to study key questions about the origin of jet activity in radio-quiet AGNs\footnote{Radio-quiet AGNs display no or weak radio emission.}, possibly triggered by minor mergers, at high physical resolution. 

One key question is the physical nature of the leading (with respect to stellar arms) radio continuum structures in NGC~7479, which sets requirements for the nuclear engine to generate the required speeds, ejection directions, and the precession mechanism. Precession may be driven by the accretion disk and the Lense--Thirring effect \citep[e.g.,][]{lu1990,merritt2002,britzen2023}, by the radiation pressure \citep[e.g.,][]{maloney1996}, or by a binary black hole resulting from a merger \citep[e.g.,][]{begelman1980,britzen2023}. 

Several observations demonstrate AGN (Seyfert) activity in the nucleus of NGC~7479. The detection of a broad H$\alpha$ line with no broad H$\beta$ line implies a Seyfert 1.9 classification \citep{ho1997}. \citet{fadda2021} report that the nuclear X-ray spectrum extracted from previously unpublished Chandra observations is typical of an AGN and that NGC~7479 harbors a heavily obscured active nucleus. A central, heavily obscured, or Compton-thick, hard X-ray source with an intrinsic luminosity $L_{\rm X,hard}$ $\approx$~10$^{42\text{--}44}$ erg~sec$^{-1}$ has been detected with Nuclear Spectroscopic Telescope Array (NuSTAR) observations \citep{marchesi2019,zhao2021,fischer2021,pizzetti2022,tanimoto2022}.

Furthermore, the compactness of the nuclear source (size less than 15~pc, see the Multi-Element Radio Linked Interferometer Network or MERLIN observations reported in this paper), and the existence of a water maser source \citep{braatz2008} point to Seyfert activity. The nuclear SMBH mass has been estimated to be 5--41$\times$10$^{6}$ $M_\sun$ \citep{fischer2021,gliozzi2024} using indirect methods. The AGN bolometric luminosity has been estimated as 4.5--8.2$\times$10$^{43}$ ergs/s \citep{pizzetti2022}. However, we note that due to the heavy extinction in the nucleus  \citep[e.g.,][]{laine1999b,fischer2021,pizzetti2022}, the mass and luminosity estimates may suffer from considerable uncertainties, and no definite conclusions can be drawn about the nature of the nucleus based on these estimates.  

There are also clear signs of ongoing starburst activity, indicated, for example, by the narrow optical emission-line ratios \citep{devereux1989}, and a soft X-ray emission component \citep{panessa2006}.
However, \citet{soifer2004} argue, based on high spatial resolution Keck Long Wavelength Specrometer observations at 12.5 $\mu$m, that a nuclear starburst does not contribute significantly to the infrared luminosity of the nucleus. A similar conclusion was also reached by \citet{fischer2021}, who showed that the estimated nuclear star formation and supernova rates make a negligible contribution to the detected nuclear radio continuum emission.

The current paper investigates the structure and excitation of the nuclear radio continuum emission in NGC~7479 with new VLBI observations.

\section{Observations and Data Reduction} \label{obs}

VLBI observations of the target source NGC~7479 were carried out with the European VLBI Network (EVN) in 2014 and with the Very Long Baseline Array (VLBA) of the National Radio Astronomy Observatory (NRAO) in 2024 and 2025. In the following, we provide the details of these observations.

\subsection{EVN}

The EVN observations of NGC~7479 were carried out at 1.7\,GHz (18\,cm, L-band) on 2014 June 2 and at 5\,GHz (6\,cm, C-band) on 2014 June 10 under project code EL048. The L-band observations included 11 EVN stations, (Effelsberg, Westerbork, Jodrell Bank, Onsala, Noto, Torun, Svetloye, Zelenchukskaya, Badary, Sheshan, and Hartebeesthoek). The C-band observations included 12 EVN stations; the above-listed stations plus Yebes.

Each frequency band was observed in a single 12\,hr session, and included observations of the fringe finder and bandpass calibrator 3C454.3, and the two phase calibrators J2300+1037 and J2310+1055, in addition to the scans on the target source. The on-target-source time was 7.2\,h per observing session.

At both frequency bands, eight 16\,MHz data channel pairs were recorded at each station, resulting in a total bandwidth of 128\,MHz per polarization. The data were correlated at the Joint Institute for VLBI in Europe (JIVE), delivering 64 spectral points per 16\,MHz subband pair (250\,kHz channel spacing).

These EVN observations utilized nodding-style phase referencing. The phase calibrators J2300+1037 and J2310+1055 are separated from the target source by $2\fdg 04$ and $1\fdg 95$, respectively. The positional uncertainties are 0.12\,milliarcseconds (mas) in right ascension (R.A.) and 0.25\,mas in declination (Dec.) for J2300+1037, and 0.04\,mas in both R.A. and Dec. for J2310+1055. Images of these phase calibrators are shown in Appendix~\ref{appendixa}.

Data reduction and analysis were performed using the Astronomical Image Processing System (AIPS; \citealt{greisen2003})
following standard VLBI data reduction procedures. The phase reference sources were self-calibrated in phase first, then in both amplitude and phase, and the solutions were applied to the target source. Deconvolution and imaging were performed using a grid weighting near the mid-point between pure natural and pure uniform weighting (ROBUST=1 in AIPS task IMAGR).

\subsection{VLBA}

The VLBA observations (in project BL318) of NGC~7479 were carried out at 1.5\,GHz (20 cm, L-band) in two 5\,h sessions on 2024 December 6 and 13, and at 4.9\,GHz (6 cm, C-band), also in two 5\,h sessions, on 2025 January 11 and 14. In addition to the ten standard VLBA stations, a single VLA station (Y1) was also included in all the observing sessions.

At L-band, eight 32\,MHz data channel pairs were recorded at each station using the ROACH Digital Backend and the polyphase filterbank (PFB) digital signal-processing algorithm, both with right- and left-hand circular polarizations, and sampled at two bits. The total bandwidth was 256\,MHz centered at 1.525\,GHz. The total on-target observing time from the two observing sessions was 6\,h.

At C-band, four 128\,MHz data channel pairs were recorded at each station using the ROACH Digital Backend and the digital downconverter (DDC) signal-processing algorithm, also both with right- and left-hand circular polarizations, and sampled at two bits. The total bandwidth was 512\,MHz centered at 4.868\,GHz. The total on-target observing time from the two observing sessions was also 6\,hr.

At both L- and C-bands, the VLBA (+Y1) observations utilized nodding-style phase referencing using the calibrator J2310$+$1055 that is separated by $1\rlap{.}{^\circ}95$ from the target source. The phase referencing cycle time was 4\,min: 3\,min on the target and 1\,min on the calibrator. Images of the phase calibrator are shown in Appendix~\ref{appendixa}.

The observations also included the calibrator source 3C\,454.3 that was used as a fringe finder and bandpass calibrator, and J2300$+$1037 as a phase-check source. The latter is separated by $2\fdg 5$ from the phase calibrator J2310$+$1055.

The data were correlated with the VLBA Distributed FX (DiFX) correlator \citep{DEL11} in Socorro, NM, with 1\,s correlator integration time. 

Data reduction and analysis were performed using AIPS and following standard VLBI data reduction procedures. Amplitude calibration was performed using measurements of the antenna gain and the system temperature of each station. The phase reference sources were self-calibrated in phase first, then in both amplitude and phase, and the solutions were applied onto the target source as well as the phase-check source. Deconvolution and imaging were performed using a grid weighting near the mid-point between pure natural and pure uniform weighting (ROBUST=1 in AIPS task IMAGR).

Examining the resulting images of the phase-check source showed an apparent shift in its position compared to the known coordinates in the VLBA calibrator database. At L-band, the peak of this source is shifted by 1.2\,mas in R.A. (i.e., to the east) and $-$3.9\,mas in Dec. (i.e., to the south). At C-band, the peak of the phase-check source is shifted by less than 0.01\,mas in R.A. and $-$1.1\,mas in Dec. These translate to a relative shift in the position of this source between L- and C- bands of $\Delta$R.A. = 1.2\,mas and $\Delta$Dec. = $-$2.8\,mas.

\section{Results} \label{results}

\subsection{EVN Observations} \label{evnresults}

\subsubsection{Morphology} \label{EVNmorphology}

\begin{figure*}[t]
  \begin{center}
    \includegraphics[width=0.49\linewidth]{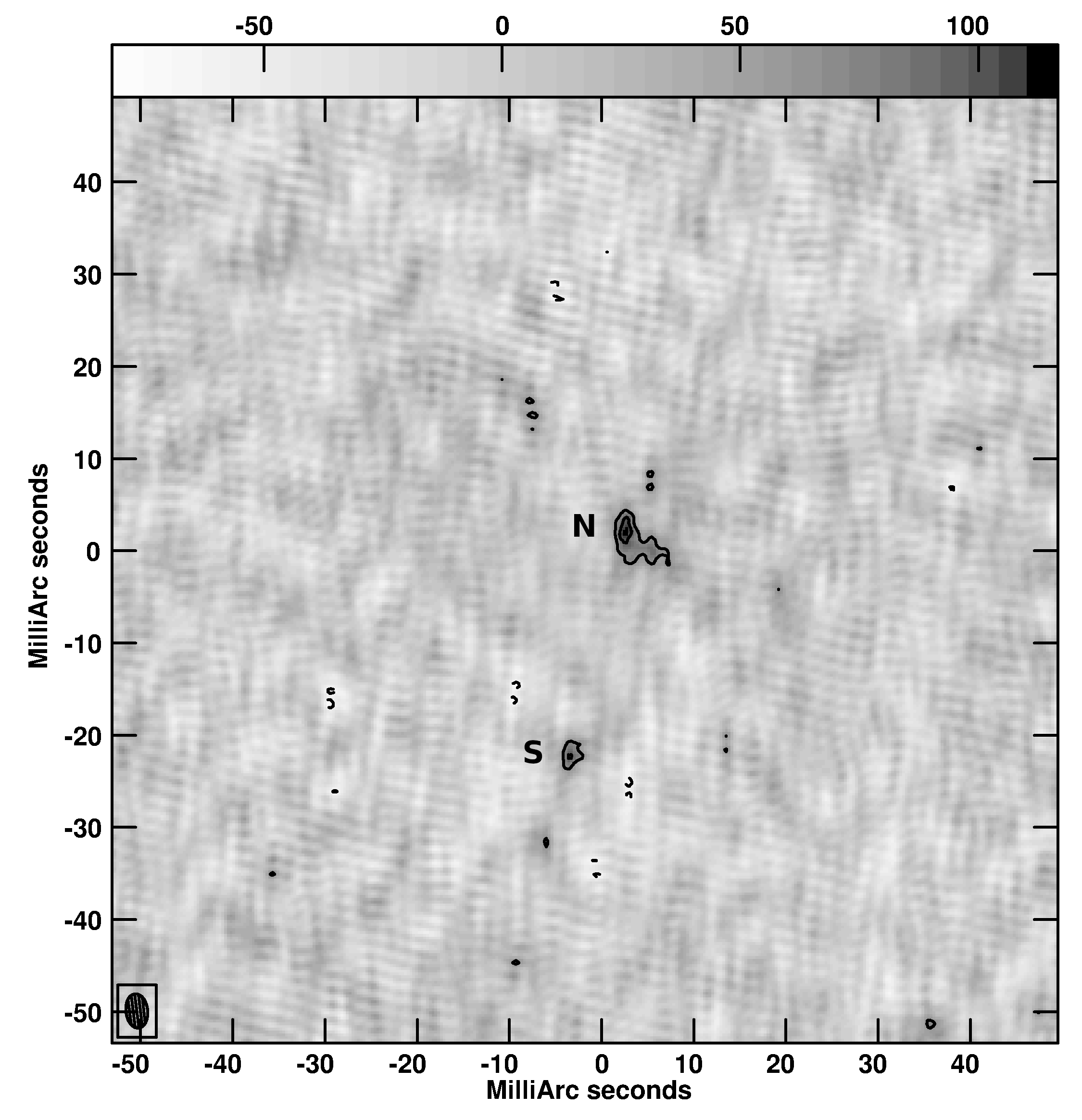}
    \includegraphics[width=0.49\linewidth]{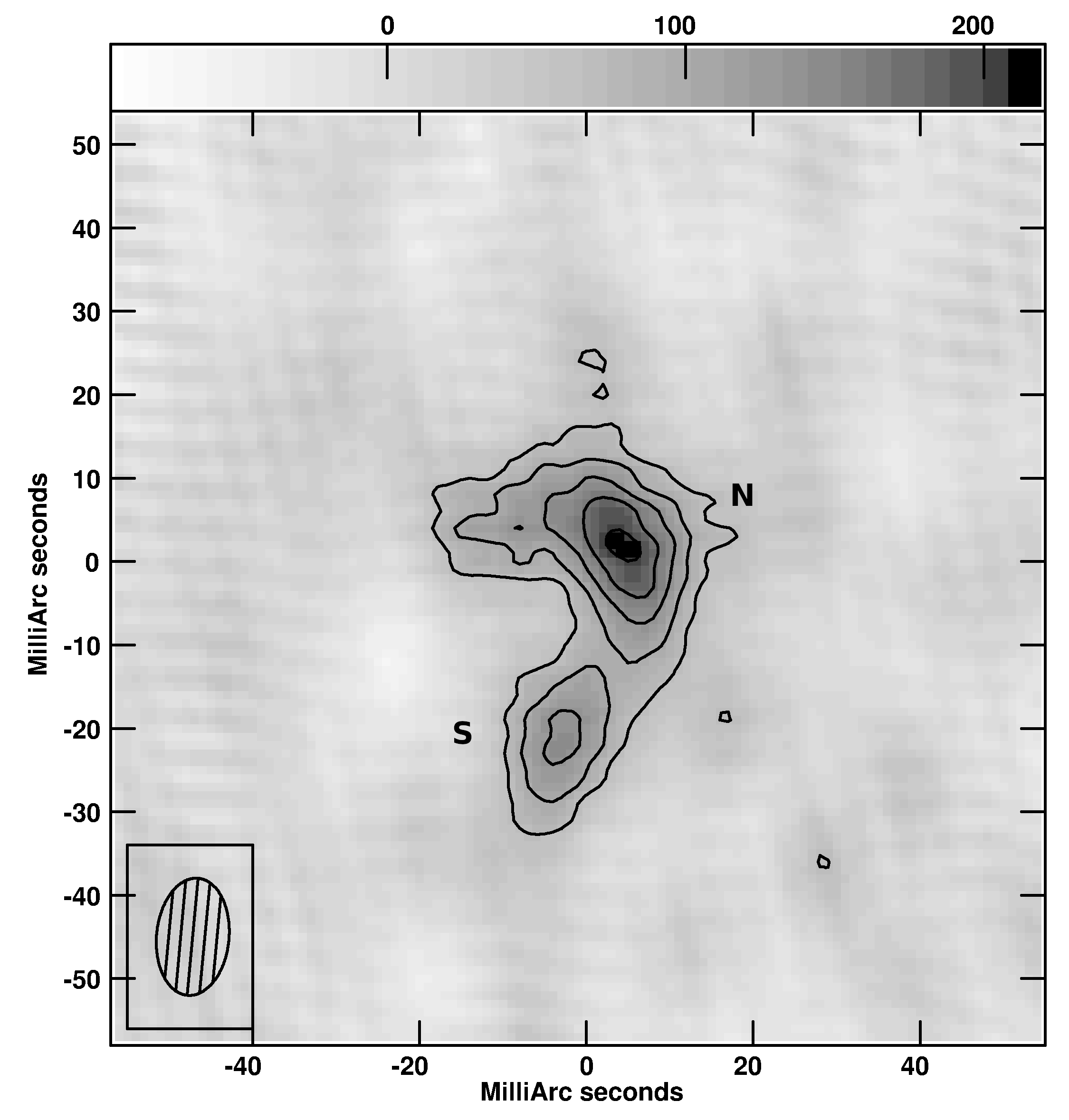}
  \end{center}
  \caption{\label{evn6and18cmfigs}6~cm (left) and 18~cm (right) EVN images of the nuclear region of NGC~7479. The contour levels are at (-4, 4, 6, 8) $\times$ 16~$\mu$Jy~beam$^{-1}$ (left) and at (-4, 4, 6, 8, 10, 12) $\times$ 17~$\mu$Jy~beam$^{-1}$ (right). The synthesized beam sizes are 3.7 mas $\times$ 2.4 mas (P.A. $=-7^{\circ}$),
  and 14.1 mas $\times$ 9.2 mas (P.A. $= 5^{\circ}$), for the 6 and 18 cm images, respectively, and are shown at the bottom left corner of each image. The gray-scale wedges are in units of $\mu$Jy~beam$^{-1}$. The letters ``N'' and ``S'' are used to label the components detected in the VLBA 6 cm image (Figure~\ref{vlba6and18cmfigs}). Note that the nominal phase center of the observations is not in the center of the images in Figures~\ref{evn6and18cmfigs}--\ref{extended-emissionfig}.}
\end{figure*}

\begin{figure*}[t]
  \begin{center}
    \includegraphics[width=0.49\linewidth]{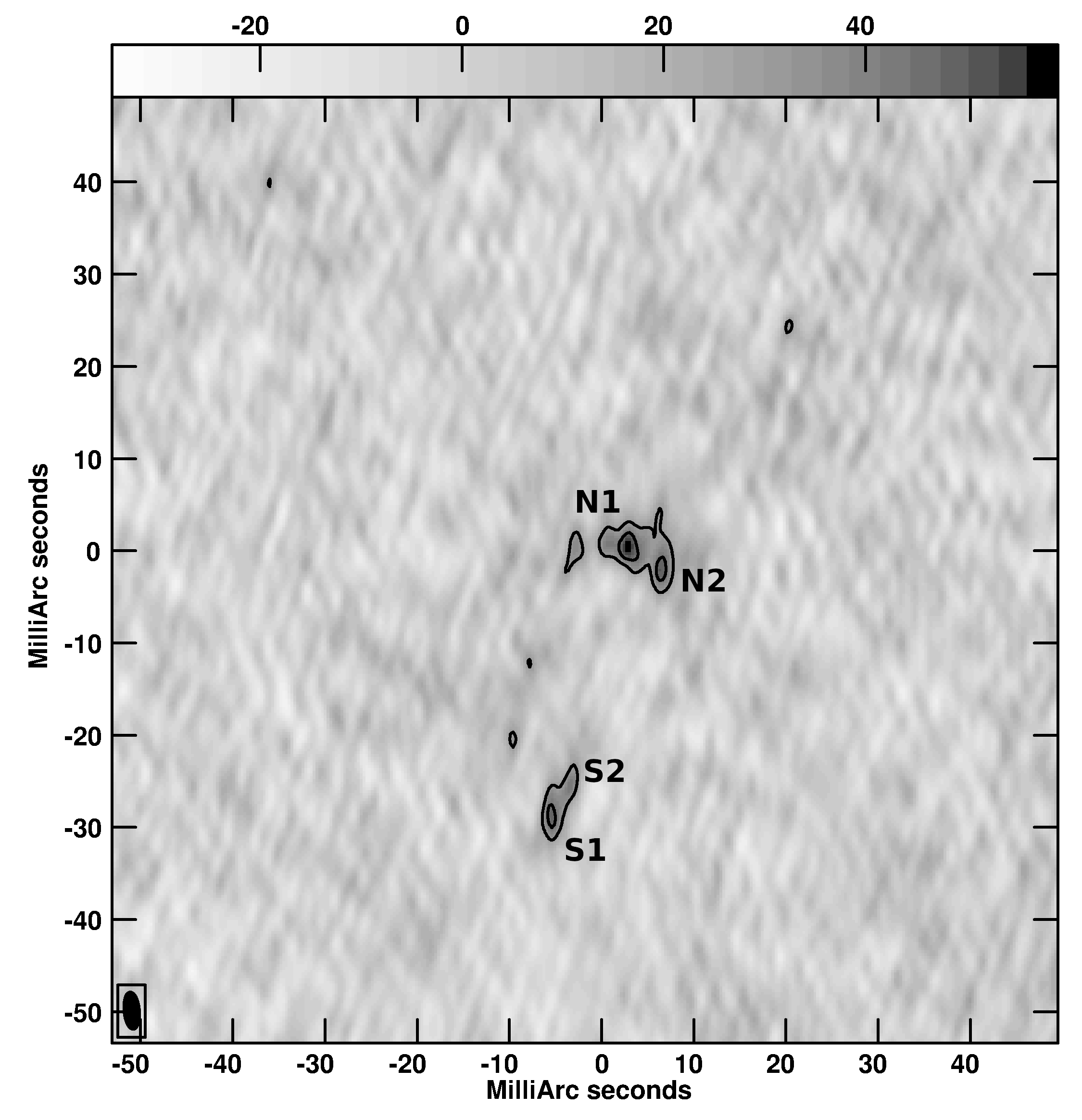}
    \includegraphics[width=0.49\linewidth]{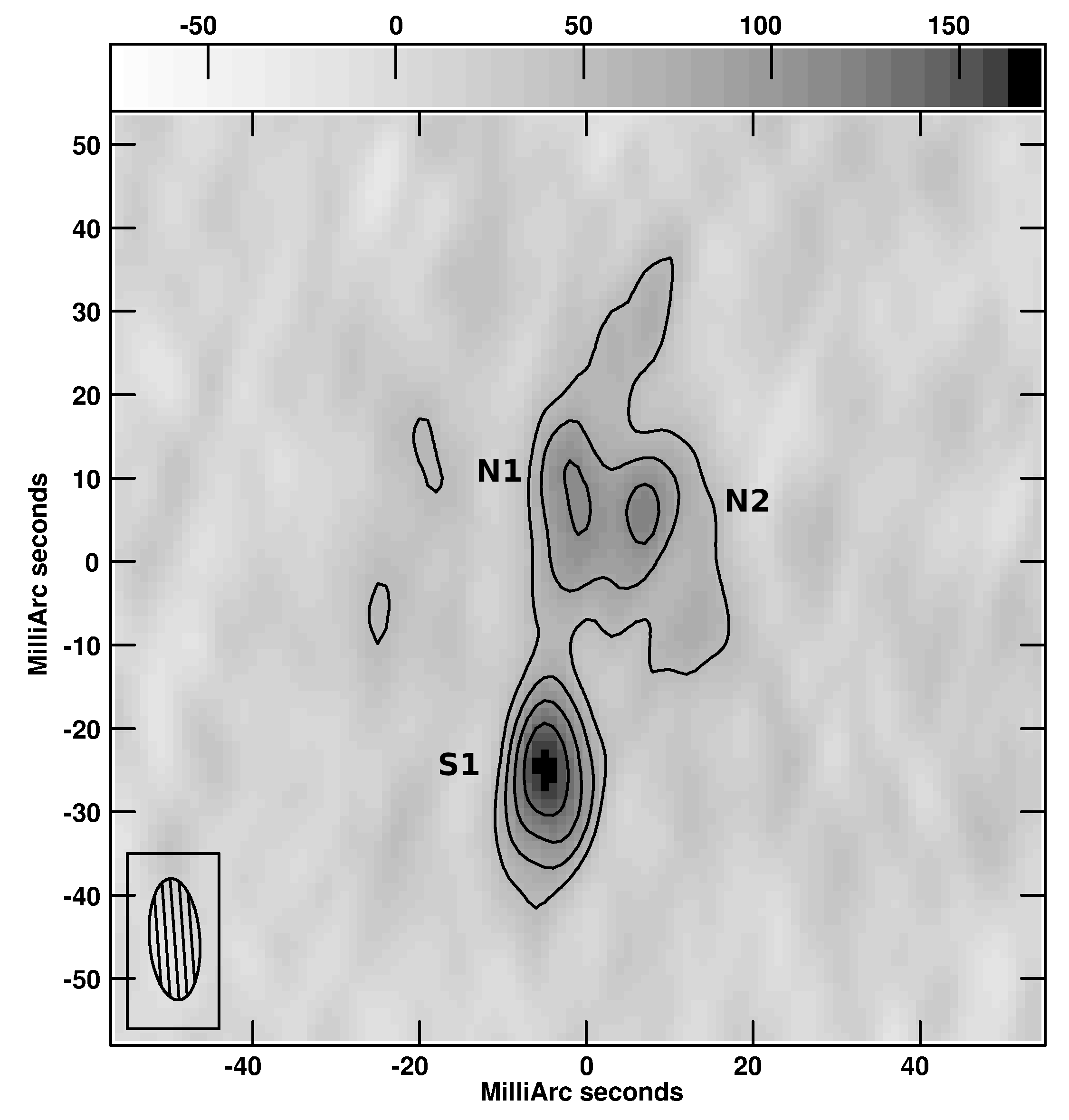}
  \end{center}
  \caption{\label{vlba6and18cmfigs}6~cm (left) and 20~cm (right) VLBA images of the nuclear region of NGC~7479. The contour levels are at (-4, 4, 6, 8) $\times$ 7.5~$\mu$Jy~beam$^{-1}$ (left) and at (-4, 4, 6, 8, 10) $\times$ 14~$\mu$Jy~beam$^{-1}$ (right). The synthesized beam sizes are 4.0 mas $\times$ 1.7 mas (P.A. $=-5^{\circ}$),
  and 14.6 mas $\times$ 6.1 mas (P.A. $= -4^{\circ}$), for the 6 and 20 cm images, respectively, and are shown at the bottom left corner of each image. The gray-scale wedges are in units of $\mu$Jy~beam$^{-1}$. The letters ``N'' and ``S'' and the numbers ``1'' and ``2'' are the labels of the radio emission components detected in the VLBA 6 cm image.}
\end{figure*}

Two emission regions (labeled ``S'' and ``N'' in Figures~\ref{evn6and18cmfigs} and \ref{vlba6and18cmfigs}) are detected at both the 6 cm and 18 cm images. These are offset from each other by about 23 mas in a south-southeast--north-northwest direction. These emission regions appear to have been marginally resolved, except for ``S'' at 6 cm, which is unresolved (see Table~\ref{tab:ngc7479measurements}). In addition, in the 18~cm image, the north-northwestern region ``N'' has an extension towards east at its northern end, with possibly another emission region about 13.5 mas east of the peak of the north-northwestern component (see components ``N1'' and ``N2'' in Fig.~\ref{vlba6and18cmfigs}).

\begin{figure}[htbp]
  \begin{center}
    \includegraphics[width=0.99\linewidth]{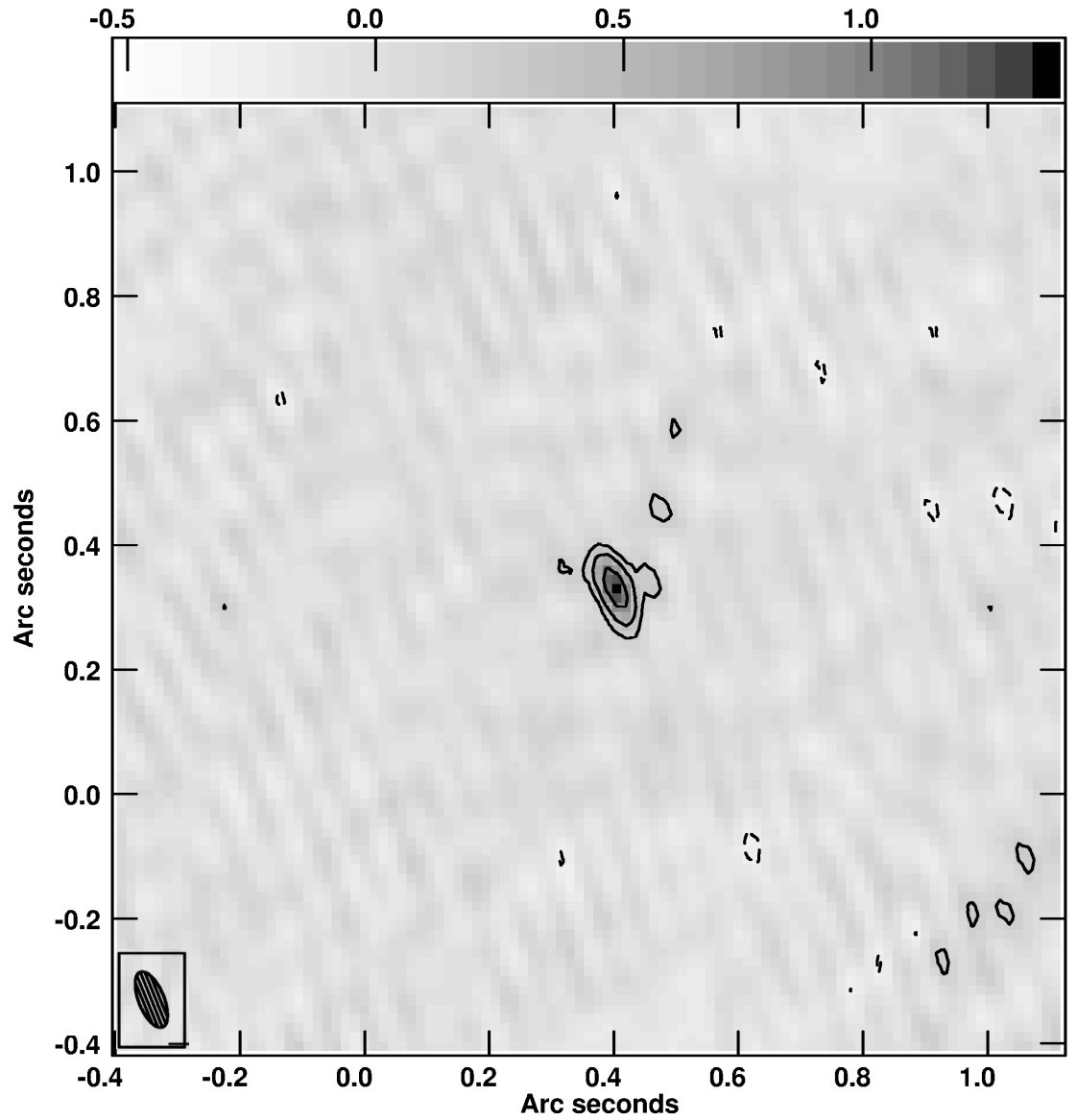}
  \end{center}
  \caption{\label{merlinfig} MERLIN 6~cm image of the nuclear region of NGC~7479. The contour levels are at (--4, 4, 6, 8) $\times$ 80~$\mu$Jy~beam$^{-1}$. The beam is at the bottom left corner.}
\end{figure}

\subsubsection{Flux densities and spectral indices}\label{EVNfluxspindex} 

Flux density measurements were performed using the Common Astronomy Software Applications for Radio Astronomy \citep[CASA;][]{casa2022} version 4.7.2. We fitted each detected emission region using a 2D Gaussian to obtain the central coordinates and the peak flux density, and its uncertainty. In cases of extended emission regions, we measured the emission inside the 3$\sigma$ contour ($\sigma$ = rms noise outside the emission regions), and estimated its uncertainty by multiplying the rms by the square root of the emitting region area expressed in beams.

The total flux density in the ``N'' source is 0.2 mJy at 6 cm and 0.7 mJy at 18 cm in the EVN images (Table~\ref{tab:ngc7479measurements}). The spectral index $\alpha$ ($S$ $\propto$ $\nu$$^{\alpha}$, where $S$ is the flux density and $\nu$ is the frequency of the emission) is $>$~$-1.1$, as some of the emission is probably resolved out in the 6 cm observations.

For the south-southeastern source ``S,'' we measured 0.08 mJy at 6 cm and 0.24 mJy at 18 cm. The spectral index is $>$~$-1.0$.

\subsubsection{Brightness temperatures}\label{EVNbrighttemp}

Brightness temperatures in Kelvin, at rest frame, were calculated from

\begin{equation}
    T_{\mathrm{B}} = 1.8\times10^9 (1+z) \frac{S_{\nu}}{\nu^2 \phi_{\mathrm{maj}}\phi_{\mathrm{min}}}~[\mathrm{K}],
\end{equation}

\noindent where $z$ is the redshift, $S_{\nu}$ the integrated flux density in mJy from the 2D Gaussian fit at the observing frequency $\nu$ expressed in GHz, and the $\phi_{\mathrm{maj}}$ and $\phi_{\mathrm{min}}$ are the fitted FWHMs of the major and minor axes of the emission region in mas, respectively, deconvolved from the beam. In case the source was unresolved, we used half the beam size as the upper limits for the source size (that is, $\phi < \theta/2$ where $\theta$ is the beam size).

For the ``N'' component, the brightness temperatures are 2.7$\times$10$^{6}$ K and 1.7$\times$10$^{6}$ K for the 6 and 18 cm observations, respectively. For the ``S'' component, they are 4.5$\times$10$^{6}$ K and 0.9$\times$10$^{6}$ K for the 6 and 18 cm observations, respectively (see Table~\ref{tab:ngc7479measurements}).

\subsection{VLBA Observations} \label{VLBAresults}

\subsubsection{Morphology} \label{VLBAmorphology}

The VLBA observations resolve the ``N'' component into two separate components, separated by about 6 mas, ``N1'' and ``N2,'' at both 6 and 20 cm (Figure~\ref{vlba6and18cmfigs}). These are separated by about 30 mas from the "S" component.

The southern component ``S'' is unresolved at 20 cm, but is resolved into two components at 6 cm. There is a hint of an eastern extension of the ``N'' component at 6 cm, similar to that seen in the EVN 18 cm observations. All the components seen with the VLBA appear unresolved, except for the ``N1'' component at 6 cm that appears extended in the east--west direction. The separation of the two components seen in north, ``N1'' and ``N2,'' is also different between the 6 and 20 cm observations, reducing to about 3 mas at 6 cm.

\subsubsection{Flux densities and spectral indices}\label{VLBAfluxspindex}

The ``N'' component resolves into two components that are not located at the same relative positions in the VLBA 6 cm and 20 cm images. The ``N1'' and ``N2'' components have flux densities of 57 $\mu$Jy and 50 $\mu$Jy at 6 cm, and 118 $\mu$Jy and 121 $\mu$Jy at 20 cm, respectively. The total flux density of the ``N'' component is 285 $\mu$Jy and 669 $\mu$Jy at 6 cm and 20 cm (Table~\ref{tab:ngc7479measurements}), respectively. The spectral index for the ``N'' component is $>$ $-0.77$. 

The ``S'' component is resolved into two components at 6 cm, ``S1'' and ``S2.'' Their flux densities are 49 and 42 $\mu$Jy. The total ``S'' component has a 6 cm flux density of 113 $\mu$Jy (see Table~\ref{tab:ngc7479measurements}). The 20 cm flux density for the total ``S'' component is 336 $\mu$Jy. The spectral index for the total ``S'' component is $>$ $-1$.

\subsubsection{Brightness temperatures}\label{VLBAbrighttemp}

For the ``N1'' and ``N2'' components, the VLBA brightness temperatures are 1.4$\times$10$^{6}$ K and 1.7$\times$10$^{6}$ K at 6 and 20 cm, respectively. For the ``S1'' and ``S2'' components at 6 cm, the brightness temperatures are 0.9$\times$10$^{6}$ K and 1.0$\times$10$^{6}$ K. At 20 cm, the ``S'' component has a brightness temperature of 2.3$\times$10$^{6}$ K (see Table~\ref{tab:ngc7479measurements}).

\movetabledown=45mm
\begin{table*}[p]
\caption{Measured properties in the EVN, VLBA, and MERLIN data.\label{tab:ngc7479measurements}}
\centering
\footnotesize
\begin{rotatetable}
\begin{tabular}{lcccccccccc}
\hline\hline                                                                       \\
Telescope  & Freq.  & Rms                 & Comp.  & $S_{\mathrm{peak}}$ & $S_{\mathrm{int}}$ & Beam size          & Beam PA  & R.A.               & Dec              & $T_{\mathrm{B}}$ \\
       & (GHz) & ($\mu$Jy bm$^{-1}$) &       & ($\mu$Jy bm$^{-1}$) & ($\mu$Jy)          & (mas)              & (deg)    & (hh:mm:ss.s)     & (dd:mm:ss.s)     & (10$^6$ K)              \\  \hline\hline
VLBA   & 1.5   & 14.0                & ``S''    & 166.1 $\pm$ 7.1     & 336.4 $\pm$ 26.9   & 14.6 $\times$ 6.0  & $-$4.1   & 23:04:56.6323110 & 12:19:22.6182610 &  2.3      \\
       &       &                     & ``N1''    & 117.6 $\pm$ 3.4     &                    &                    &          & 23:04:56.6320814 & 12:19:22.6504521 &  1.4       \\
       &       &                     & ``N2''    & 121.2 $\pm$ 1.8     &                    &                    &          & 23:04:56.6315420 & 12:19:22.6497586 &  1.7      \\
       &       &                     & ``N'' &                     & 668.6 $\pm$ 43.1   &                    &          &                  &                  &         \\ \hline
VLBA   & 4.9   & 7.5                 & ``S1''    & 49.4 $\pm$ 1.9      & 113.2 $\pm$ 13.8   & 4.0 $\times$ 1.6   & $-$5.2   & 23:04:56.6323631 & 12:19:22.6153877 &  0.9        \\
       &       &                     & ``S2''  & 41.6 $\pm$ 1.5      &                    &                    &          & 23:04:56.6322460 & 12:19:22.6183116 &  1.0      \\    
       &       &                     & ``N1''    & 56.8 $\pm$ 1.9      &                    &                    &          & 23:04:56.6318028 & 12:19:22.6444123 &  1.4      \\
       &       &                     & ``N2''    & 49.8 $\pm$ 1.1      &                    &                    &          & 23:04:56.6315576 & 12:19:22.6421053 &  0.9     \\
       &       &                     & ``N'' &                     & 285.3 $\pm$ 22.0   &                    &          &                  &                  &          \\ \hline
EVN    & 1.7   & 17.0                & ``S''    & 139.8 $\pm$ 9.1     & 235.5 $\pm$ 27.7   & 14.1 $\times$ 8.7  & 5.25     & 23:04:56.632127  & 12:19:22.6242440 &  0.9      \\
       &       &                     & ``N'' & 200.0 $\pm$ 14.0    & 668.1 $\pm$ 45.2   &                    &          & 23:04:56.631757  & 12:19:22.6463690 &  1.7       \\ \hline
EVN    & 5.0   & 16.0                & ``S''    & 84.0 $\pm$ 13.0     & point source       & 3.7 $\times$ 2.3   & $-$6.74  & 23:04:56.632214  & 12:19:22.6219320 &  4.5        \\  
       &       &                     & ``N'' & 109.4 $\pm$ 7.2     & 236.7 $\pm$ 30.2   &                    &          & 23:04:56.631819  & 12:19:22.6459378 &  2.7      \\ \hline
Merlin & 5.0   & 80.0                &       & 1282.5 $\pm$ 80     & point source       & 95.2 $\times$ 40.5 & 22.3     &                  &                  &        \\  \hline 
\end{tabular}
\end{rotatetable}
\end{table*}

\begin{figure}[htbp]
  \begin{center}
    \includegraphics[width=0.99\linewidth,trim={1cm 2cm 0.6cm 1.3cm},clip]{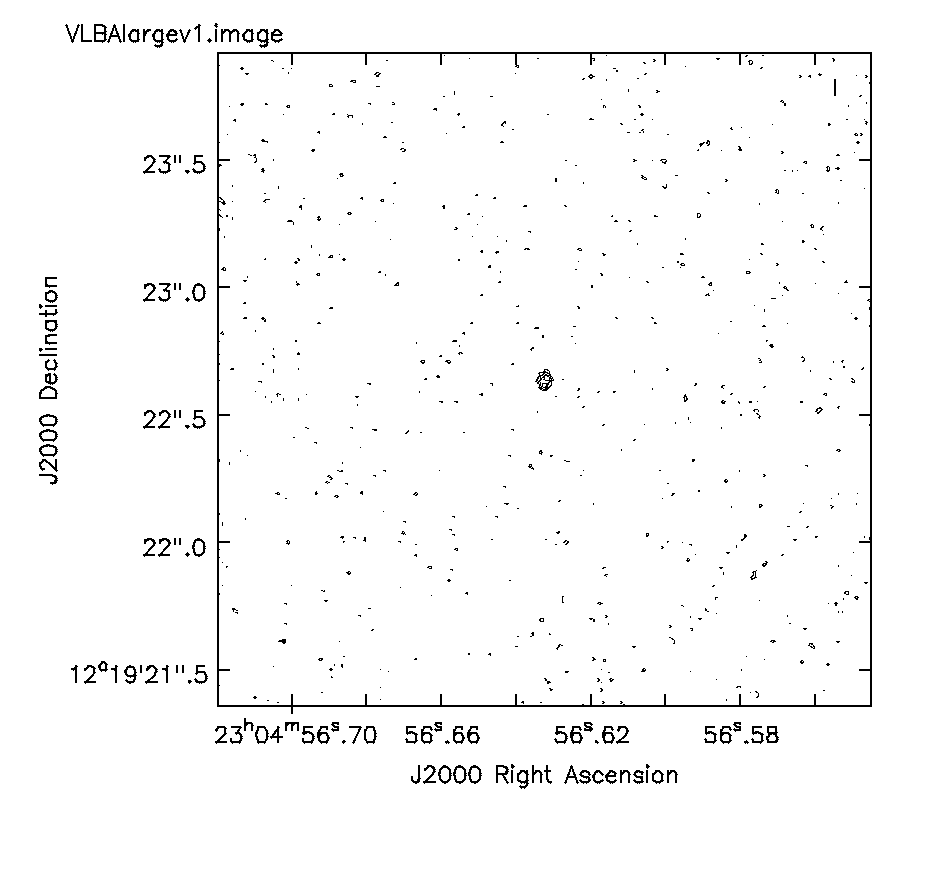}
  \end{center}
  \caption{\label{extended-emissionfig}\small VLBA 20~cm large area image produced using natural weighting. This image shows an area out to about 1\farcs 3 radius around the nucleus of NGC~7479 and reveals the lack of any detectable continuous extended emission outside the nucleus. The contour levels are at 12~$\mu$Jy beam$^{-1}$ $\times$ (-3, 3, 6, 12). The beam size is 0\farcs018 $\times$ x 0\farcs016, and the PA is 16.6 deg. The beam is near the lower left corner but unresolved because of its small size.}
\end{figure}

\subsection{MERLIN Data}
\label{merlindata}
We also present 6~cm MERLIN imaging taken on 1998 February 8 and reduced following standard reduction procedures (Figure~\ref{merlinfig}). The flux density of the unresolved nuclear source in these 0\farcs 1$\times$0\farcs 04 beam size observations is 1.3~mJy. We made a $uv$-tapered image as well, but it did not reveal any extended emission either. The unresolved ``core'' morphology is perhaps most similar to the LINER-type morphology in the LeMMINGs survey that was taken with the $e$-Merlin array at 1.4 and 5 GHz \citep{baldi2018,baldi2021a,baldi2021b,williams2026}. While the LeMMINGs studies have shown that 
nearby nuclear starburst galaxies may also have a radio core at 
MERLIN-detectable resolution, consistent with what we see 
in Figure~\ref{merlinfig}, they are extremely rare. Based on the evidence presented in Sections~\ref{intro} and \ref{discussion:origin}, we believe that the contribution from star
formation to the detected radio continuum flux density is 
negligible.

\subsection{Large Field Observations}
\label{largefield}

We have also imaged a much larger field around the pointing center in the VLBA observations at 20 cm. Figure~\ref{extended-emissionfig} shows the large field of the VLBA, cleaned using natural weighting. Extended emission areas above 3$\sigma$ of the noise level are not detected outside the central sources discussed above, out to about 1 arcsecond radius from the nucleus. We tried increasing the sensitivity to extended emission by using a $uv$-taper, resulting in a beam about three times the size of the natural weighted clean beam, but this also failed to bring out any extended emission.

\subsection{Relative Motion of Detected Emission Regions}
\label{movement}
As we have not identified a nucleus in a definite manner among the detected emission regions, we cannot compare the two epochs of VLBI data by fixing the position of one emission region and seeing how much the distances between it and other emission regions have changed. Nor can we use the recorded coordinates as absolute correct positions because of the relatively large uncertainties in the coordinates due to the uncertainties in phase referencing. The only thing we can do is to measure the relative distances between the emission regions in the two observed bands and on the given (averaged) observation dates in the images and assume that the two emission regions ("N" and "S") correspond to the same ones between the 2014 and 2024/2025 epochs.

We calculate the uncertainty in the relative distances ($\Delta\theta$) of emission regions using Equation (1) from \citet{goddi2004}:

\begin{equation}
\label{eq2}
\Delta\theta = \frac{\sigma}{2I}FWHM,
\end{equation}

\noindent where $\sigma$ is the noise off-source in rms, $I$ is the peak intensity, and {\it FWHM} is the size of the unconvolved region.

\begin{deluxetable*}{lcccc}
\tablewidth{0pt}
\tablecaption{Measurements of the distance between ``N'' and ``S'' emission regions. \label{table:distances}}
\tablehead{
\colhead{Array} & \colhead{Date} & \colhead{Freq.} & \colhead{Distance} & \colhead{Unc.} \\
\colhead{} & \colhead{} & \colhead{(GHz)} & \colhead{(mas)} & \colhead{(mas)} }
\startdata
EVN & 2014.42 & 1.7 & 23 & 2 \\
EVN & 2014.44 & 5.0 & 25 & 2 \\
VLBA & 2024.95 & 1.5 & 33 & 2 \\
VLBA & 2024.03 & 4.9 & 30 & 1 \\
\enddata
\tablecomments{Uncertainties are based on Equation ~\ref{eq2}. For the VLBA L-band, the coordinates of the two resolved emission regions, ``N1'' and ``N2'' were averaged to get an ``N'' position. For the VLBA C-band, we used the averaged coordinates of ``N1'' and ``N2'' for an ``N'' position and the coordinates of the strong emission region ``S1'' as ``S'' position.}
\end{deluxetable*}

Measuring the distance between the average positions of the ``N'' and ``S'' components and taking the difference in the distance between the EVN and VLBA observation epochs, we see values of about 5--10 mas (see Table ~\ref{table:distances} and Figure~\ref{separations}). These numbers are much larger than the individual source positional uncertainties, calculated using Equation~\ref{eq2} above that are about 2 mas individually, or less than 3 mas for the distance between the components. 

Therefore, the components appear to have moved apart by this amount, as seen in the C- and L-band images, in about ten years (or at an angular separation rate of $\mu$~$\approx~$ 0.7 mas/yr). For a distance of 32 Mpc and a redshift of $z$ = 0.007925, this yields an average apparent speed of $\beta_{\rm app}$ = 0.35c (where c is the speed of light), i.e., the apparent motion would be relativistic, but subluminal. Such speeds are very typical for jet motion detected in nearby radio galaxies (e.g., Cyg A; \citeauthor{steenbrugge2008} \citeyear{steenbrugge2008}, \citeauthor{boccardi2016} \citeyear{boccardi2016}, 3C 84; \citeauthor{foschi2025} \citeyear{foschi2025}). We discuss the implications of the apparent movement in Section~\ref{discussion:origin} below.

\begin{figure}[htbp]
  \begin{center}
    \includegraphics[width=0.99\linewidth]{FIGURES/separations.jpg}
  \end{center}
  \caption{\label{separations}\small Relative positions of the VLBA and EVN components compared to the VLBA 5~GHz S1 component. Error bars indicate the maximum positional uncertainty of each component, 2~mas. }
\end{figure}

\section{Discussion} \label{discussion}

\subsection{Nature of the Two Main Nuclear Radio Continuum Emission Regions}\label{discussion:nature}

Previous VLBI observations of low-luminosity AGNs \citep[e.g.,][]{falcke2000,ulvestad2001,nagar2002,anderson2004,filho2004,giroletti2009,bontempi2012,panessa2013,fischer2021,cheng2025} usually revealed, when sensitivity allowed, a single nuclear point-like radio source, or, in some cases, a nuclear point source together with VLBI jet components. NGC~7479 was observed earlier with the VLBA at 6~cm by \citet{fischer2021}, who took a 1-hour integration and did not detect any emission in observations centered at the position of the nucleus. Such an outcome is not surprising, as their 1-$\sigma$ noise was close to the peak level of the sources we detected with the VLBI at 6~cm. What we have found in our VLBI observations of NGC~7479 is very unusual, as here we see an extended northern source "N" (extended mostly in the northeast--southwest direction) and another emission region "S" to the south--southeast from it, extended mostly in the northwest--southeast direction. It is possible that a jet could have apparently bent by about 30$^{\circ}$ due to projection effects (comparing the orientation of the kpc-scale jet to the position angle between the ``N'' and ``S'' VLBI components detected here). 

Assessing which of the detected emission regions, if either, is the true nucleus of NGC~7479 is difficult. The brightness temperatures are all less than a few times 10$^{6}$~K, unlikely to be due to star formation, but similar to the values of the compact radio features seen in other Seyfert nuclei \citep[e.g.,][]{gallimore1997,ulvestad1998,carilli1998}.

Assuming that one of the two detected major radio emission regions is the nucleus, and the other one is a location where a nuclear outflow wind or jet encounters a thick interstellar medium, we can calculate the time required for the VLBI component to move to its current projected separation from the nucleus. Using a measured current projected distance between the two emission regions ``N'' and ``S'' of 30 mas, a distance of 32 Mpc to NGC~7479 \citep{laine1998}, and a jet speed of 0.35c, as calculated in Section~\ref{movement}, and assuming that one of the ``N'' or ``S'' emission regions is the nucleus, the minimum age of the nuclear outflow or jet to propagate the distance between the nucleus and the moving emission region is 43 years.

Another possibility is that one of the detected radio emission regions is the nucleus of a recently merged galaxy. However, this is unlikely, as according to the \citet{laine1999} minor merger model, the remnants of a recently merged companion should lie currently at about 17$\arcsec$ north of the nucleus.

Assuming that the ``S1'' component is the nucleus, components ``S2,'' ``N1,'' and ``N2'' could be emission regions of high-density gas where a jet or an AGN-driven outflow wind encounters a dense interstellar medium (see Figure~\ref{vlba6and18cmfigs}, left). Emission regions ``S1,'' ''S2,'' and ``N2'' lie along the same line, and could define one "edge" of this outflow (or wide jet), as discussed in the weak Seyfert context by \citet{fischer2021,fischer2025}. This would ``naturally'' explain the geometry of emission region ``N1'' that is elongated perpendicular to the large-scale outflow or jet. ``N1'' and ``N2'' could mark the opposite limbs of a conically expanding outflow or jet. Edge- or limb-brightened jets are often seen in VLBI observations, not only in blazars \citep[e.g.,][]{ros2020} but also in nearby galaxies, including Cen A \citep[e.g.,][]{janssen2021}, 3C84 \citep[e.g.,][]{giovannini2018}, and M87 \citep[e.g.,][]{kim2018}. In this scenario, the observed VLBI emission could be interpreted as part of a kpc-long, slightly bent jet.

Otherwise, if the ``N'' component is the nucleus, then the east--west elongation around the nucleus is surprising. If the elongation was caused by a recent merger (an earlier merger event than the minor merger previously suggested by \citeauthor{laine1999} \citeyear{laine1999}), then it would imply a strong spin-flip of the SMBH, which is perhaps unlikely in the gas-rich nuclear environment that would make the spins align. Also, the compactness of the ``N'' region would then likely imply a recent minor merger, but there are no signs of other recent minor mergers in this system (in this scenario, the ``S'' components would be part of an older jet, and the ``N'' components would reflect the current jet direction). 

It is also possible that we did not detect the true nucleus of the galaxy, and we only see a few off-nuclear starburst regions. The emission could be dominated by optically thin synchrotron emission from cosmic rays accelerated by supernovae, consistent with the measured spectral index values. The luminosities at 6 cm are similar to those of supernova remnants at age 50--100 years \citep[e.g.,][]{colina1993,fenech2010}. Unfortunately, at the distance of NGC~7479 the current X-ray observatories would be uncapable of resolving supernova remnants. The integrated flux densities of the ``N'' and ``S'' regions in the C- and L-bands appear to be fairly similar between the two epochs of observations separated by about ten years, which allows for the possibility of off-nuclear starburst regions. Also, the brightness temperatures are low enough to allow the existence of starburst regions. However, as noted earlier, the two emission regions, ``N'' and ``S,'' seem to have moved away from each other by a detectable amount, which is not expected if they were both starburst regions.

Future radio continuum observations with even higher spatial resolution, and if possible, with polarization information, would be useful in assessing the true nature of the detected nuclear emission regions.  

\subsection{Implications from MERLIN's Radio Continuum Detection}
\label{discussion:merlin}
The MERLIN observation at 6 cm, shown in Figure~\ref{merlinfig}, only reveals an unresolved point source. Therefore, drawing conclusions about the nature of a nuclear source from images with a resolution of about 0\farcs 1 is dangerous, based on what we have discovered in our VLBI images of NGC~7479. Apparently, a lot of the emission at 0\farcs 1 and 1\arcsec\ resolution observations has been resolved out in the VLBI images, as the total detected nuclear flux density in MERLIN observations at 6 cm was 1.3 mJy and the total measured flux density is only a few hundred $\mu$Jy in our 6 cm VLBI observations.

\subsection{Driving Mechanism for the Radio Continuum Emission}\label{discussion:origin}

The arguments for a low-power, jet-induced \citep[e.g.,][]{miller1993} triggering of the radio continuum emission in the nucleus of NGC~7479 include the fact that the emission regions we have detected within a few parsecs of the nucleus are aligned with the 100 pc to kiloparsec scale radio emission, as one would expect for a relativistic, collimated jet. The radio continuum emission created when a jet interacts with interstellar clouds should also be nonthermal synchrotron radiation, consistent with the measured spectral indices.

As stated earlier in Section~\ref{movement}, the apparent distance shift between emission regions ``N'' and ``S'' in about ten years implies a relativistic speed, as is observed with radio jets \citep[e.g.,][]{blandford2019}. The wavelength-dependent shifts could be due to the so-called core shift effect, which implies a positional displacement in synchrotron self-absorbed emission regions, caused by different opacities within the regions ``S'' and ``N'' \citep[e.g.,][]{sokolovsky2011}.

The emission regions are also compact (making them VLBI-detectable), as would be expected when a collimated jet impacts the interstellar medium. The brightness temperatures are also high, $>$10$^{6}$ K, as seen in the nuclear jets of nearby low-luminosity AGNs \citep[e.g.,][]{nagar2005}.

The radio continuum spectral index $\alpha$ in relativistic shocks has a power law distribution ($\alpha = \frac{s - 1}{2}$; distribution of relativistic electrons $n(E)$$\propto$$E^{-s}$, $E$ is the energy of the electrons) where the steepness indicates the energy of the relativistic electrons, with steeper indices ($\alpha$ $<$ $-0.7$) indicating aging electrons, and flatter indices ($\alpha$ $>$ $-0.4$) suggesting younger electrons \citep[e.g.,][]{lisenfeld2010}. Since we expect that the radio continuum emission in regions within a few parsecs of the nucleus would be due to relatively recent shocks, we would have expected to see flatter spectral indices than what we observed ($\alpha$ $>$ $-0.8$). However, the observed steep spectrum could also be due to spectral aging, if the radiating electrons are not accelerated further. 

The absence of a flat spectrum nuclear core component could be due to inadequate sensitivity, or, if one of the detected emission regions is the AGN core, its spectral index could be steep, as seen in some other AGNs by, e.g., \citet{popkov2025}.

In addition to the relativistic jet scenario discussed above, the observed VLBI morphology can also be explained within a wind-driven shock framework. In this picture, the compact radio knots trace localized sites where a slower, wide-angle AGN outflow encounters dense circumnuclear gas. Such shocks naturally produce non-thermal synchrotron emission with steep spectral indices ($\alpha \leq$ –0.7), consistent with the values measured for the VLBI components. Unlike relativistic jets, wind-driven shocks are not expected to produce a bright, flat-spectrum radio core; thus, the absence of a compact nuclear source in our data does not pose a difficulty for this scenario.

A wind/shock origin also provides an alternative explanation for the apparent change in component separation between the two epochs. 
In this scenario, the observed displacement does not trace the bulk motion of a coherent plasma knot, but instead reflects the appearance of new shock sites at locations offset from previously shocked gas.
Rather than tracing the bulk motion of a long-lived jet knot, the VLBI components may represent different parcels of gas that become shock-illuminated at different times as the outflow encounters fresh dense parcels of gas. In such a case, an apparent increase in component separation can arise even if no material travels between the two locations at relativistic speeds, potentially leading to an over-interpretation of the inferred velocity, if interpreted as true proper motion. Apparent “proper motions” of this kind have been reported in other radio-quiet AGNs \citep[e.g.,][]{fischer2023}.

The observed brightness temperatures ($T_B \approx$ a few$\times$10$^6$ K) lie within the range commonly found in VLBI studies of radio-quiet AGN outflows and shocked regions in circumnuclear gas, which typically span 10$^5$--10$^8$ K \citep{shuvo2024}. These values are lower than those associated with compact jet cores \citep[$\ge$~10$^9$ K, e.g.,][]{Homan2021}.

The alignment of the VLBI components with the kpc-scale radio structure does not uniquely require a jet, as wide-angle winds channeled by the dense interstellar medium can produce collimated large-scale radio morphologies \citep[e.g.,][]{fischer2025}.

Other possibilities for producing the detected radio continuum emission include radio supernovae and supernova remnants, for which the spectral index can be in the negative range that we have observed. The likelihood of there being radio supernovae is low, as it would require a very active starburst. However, the measured radio spectral indices should be interpreted with caution because of the different spatial resolution at different wavelengths.

Finally, nuclear radio continuum emission could be free-free emission from a molecular torus, from a disc wind or the X-ray corona itself \citep{gallimore2004}. However, our measured spectral indices are much steeper than the flat or positive spectral indices expected from free--free emission.

Future spectral observations in the infrared, especially with the Near-Infrared Spectrograph (NIRSpec) and Mid-Infrared Instrument (MIRI) instruments aboard the James Webb Space Telescope (JWST), will be crucially important to determine the presence of a possible nuclear outflow wind. Such MIRI observations are currently scheduled (JWST program id 7140, PI Lee Armus). Also, intermediate spatial resolution ($0\farcs 01$--$0\farcs 1$) radio continuum observations, possibly with the Square Kilometre Array, should finally reveal the spatial distribution of the nuclear core radio emission that was previously unresolved in VLA and MERLIN intermediate-resolution observations.

\subsection{Large Area Survey}
\label{discussion:largesurvey}

There still remains a large gap in jet-like emission in NGC~7479 between the possible nuclear jet/outflow emission at the few mas scale, as reported in the previous subsections, and the large-scale jet-like structure, seen as close to the nucleus as a couple of arcseconds north \citep{ho2001,laine2006}. Our VLBA search for any extended radio emission at 20 cm, made with natural weighting and shown in Figure~\ref{extended-emissionfig}, does not reveal any signs of emission regions beyond the nucleus out to about 1 arcsecond distance. We made a $uv$-tapered image as well, with the same non-detection result for extended emission. It is possible that there is radio continuum emission at 0\farcs 03--1\arcsec\ distances from the nucleus, but it is too weak to be detected, or the outflow/jet has not encountered any dense gas in this radial range. The MERLIN image in Figure~\ref{merlinfig} with an intermediate resolution, while not very deep, does not show any definite signs of emission outside a point-like nuclear component either, within about 0.5 arcsecond radius. We made a $uv$-tapered image as well, and it did not show any extended emission either.

\section{Conclusions}
\label{conclusions}

We have conducted the highest resolution and deepest C- and L-band radio continuum observations of the nucleus of the nearby barred spiral galaxy NGC~7479 with both the EVN and the VLBA arrays with a few mas (corresponding to a few parsecs) spatial resolution. We did not detect a clear nuclear core component with a flat or positive spectral index and a brightness temperature above 10$^{8}$ K. Instead, we detected two emission regions that are marginally resolved into point-like sources with spectral indices ($\alpha$ $\approx$ $-1$) that are common to radio jet or AGN-driven wide-angle outflow wind generated shock emission. The brightness temperatures around a few times 10$^{6}$ K are indicative of synchrotron emission generated by shocks, driven by either faint and aging radio jets or a nuclear AGN wind. In addition, we see extended emission connected to these point-like emission regions, consistent with them being sub-parsec scale jets or evidence of an AGN-driven nuclear wind that consumes fresh gas.

We have detected an apparent increase in the distance between the two emission regions ``N'' and ``S,'' from the 2014 epoch to the 2024--2025 epoch, on the order of 7 mas in ten years, implying a relativistic jet speed of 0.35c. If the apparent motion is related to a jet, we estimate the age of a shock in the jet to be around 43 years, very recent.

The apparent change in the distance between the two major radio emission regions detected in our VLBI observations can also be explained by a nuclear wind that encounters fresh parcels of gas at various times.

Finally, a search for any radio continuum emission outside the nuclear region up to about 1 arcsecond radius did not produce any detections in the L-band VLBA data.

\begin{acknowledgments}

We are grateful to Drs.\ Peter Boorman and Thomas Lai and for valuable comments that substantially helped to improve the manuscript. TGP thanks Liam Owens (Craft Academy for Excellence in Science and Mathematics and Morehead State University) for his assistance with analyzing the Chandra observations of NGC 7479. This work was authored by an employee of Caltech/IPAC under Contract No. 80GSFC21R0032 with the National Aeronautics and Space Administration. The European VLBI Network is a joint facility of independent European, African, Asian, and North American radio astronomy institutes. Scientific results from data presented in this publication are derived from the following EVN project codes: AL048A and AL048B. The EVN data are partially based on observations with the 100-m telescope of the MPIfR at Effelsberg. This work made use of the DiFX software correlator developed at Swinburne University of Technology as part of the Australian Major National Research Facilities program. The National Radio Astronomy Observatory (NRAO) is a facility of the National Science Foundation operated under cooperative agreement by Associated Universities, Inc. AIPS is produced and maintained by the The National Radio Astronomy Observatory (NRAO). This research has made use of the NASA/IPAC Extragalactic Database (NED), which is operated by the Jet Propulsion Laboratory, California Institute of Technology, under contract with the National Aeronautics and Space Administration. This research has made use of NASA's Astrophysics Data System Bibliographic Services.

\end{acknowledgments}

\facilities{EVN, VLBA, MERLIN}

\software{AIPS \citep{greisen2003}}, CASA\citep{casa2022}

\newpage

\appendix

\section{Calibrator images}
\label{appendixa}

\begin{figure*}[!htb]
  \begin{center}
    \includegraphics[width=0.49\linewidth]{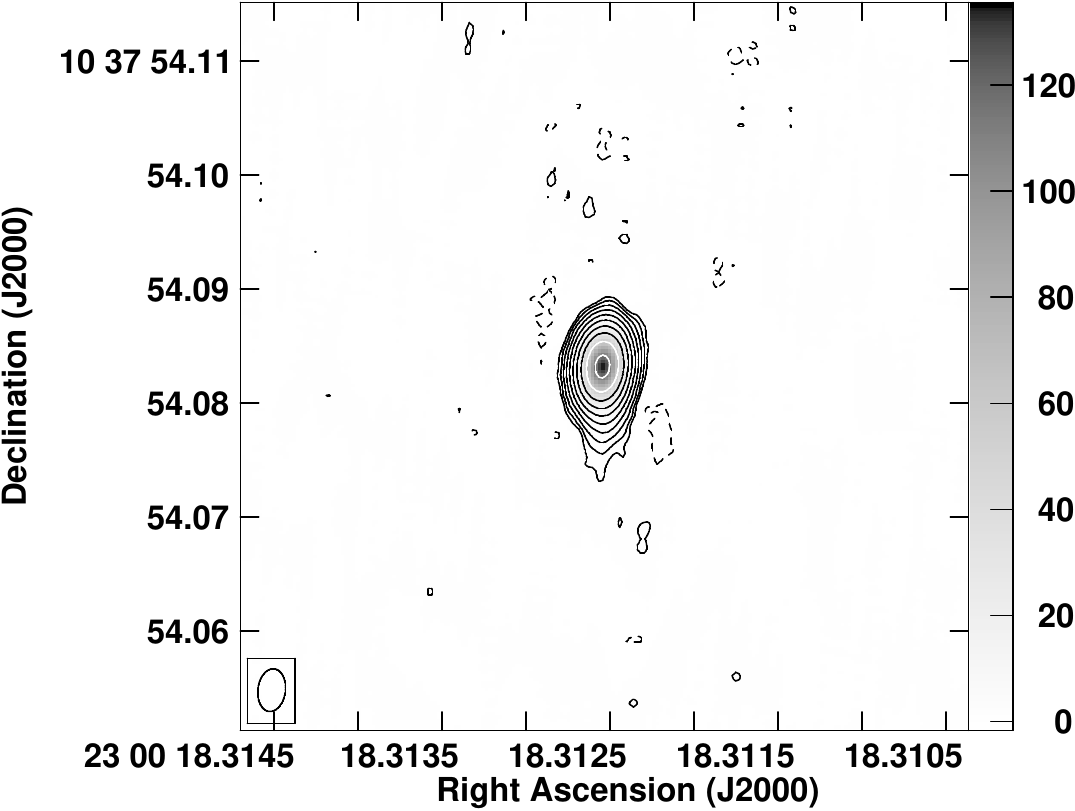}
    \includegraphics[width=0.49\linewidth]{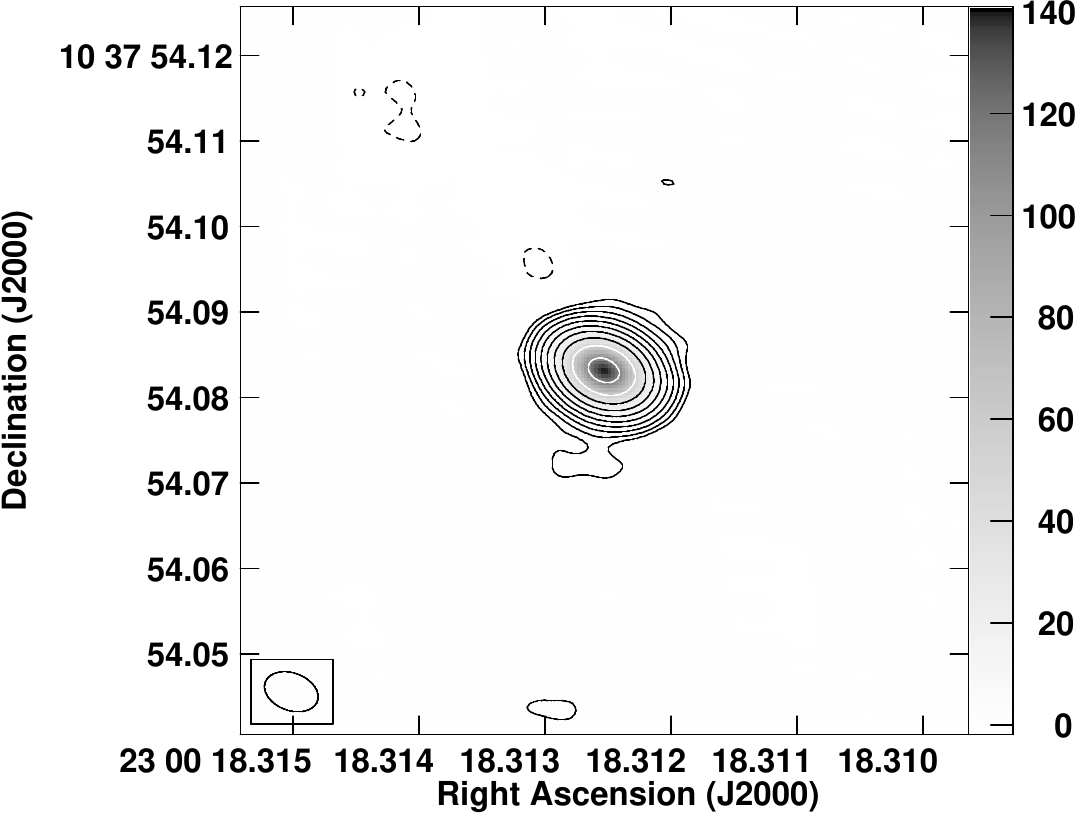}
  \end{center}
  \caption{\label{evnj2300}6~cm (left) and 18~cm (right) EVN images of the phase calibrator J2300+1037. The contour levels are at (-3, 3, 6, 12, 24, 48, 96, 192, 384, 768, 1536) $\times$ 72.83~$\mu$Jy~beam$^{-1}$ (left) and at (-3, 3, 6, 12, 24, 48, 96, 192, 384, 768, 1536) $\times$ 72.64~$\mu$Jy~beam$^{-1}$ (right). The beams are at the bottom left corners and their sizes are 3.8 mas $\times$ 2.4 mas, P.A. = -7$^{\circ}$ (6 cm, left) and 6.5 mas $\times$ 4.4 mas, P.A. = 70$^{\circ}$ (18 cm, right).}
\end{figure*}

\begin{figure*}[!htb]
  \begin{center}
    \includegraphics[width=0.49\linewidth]{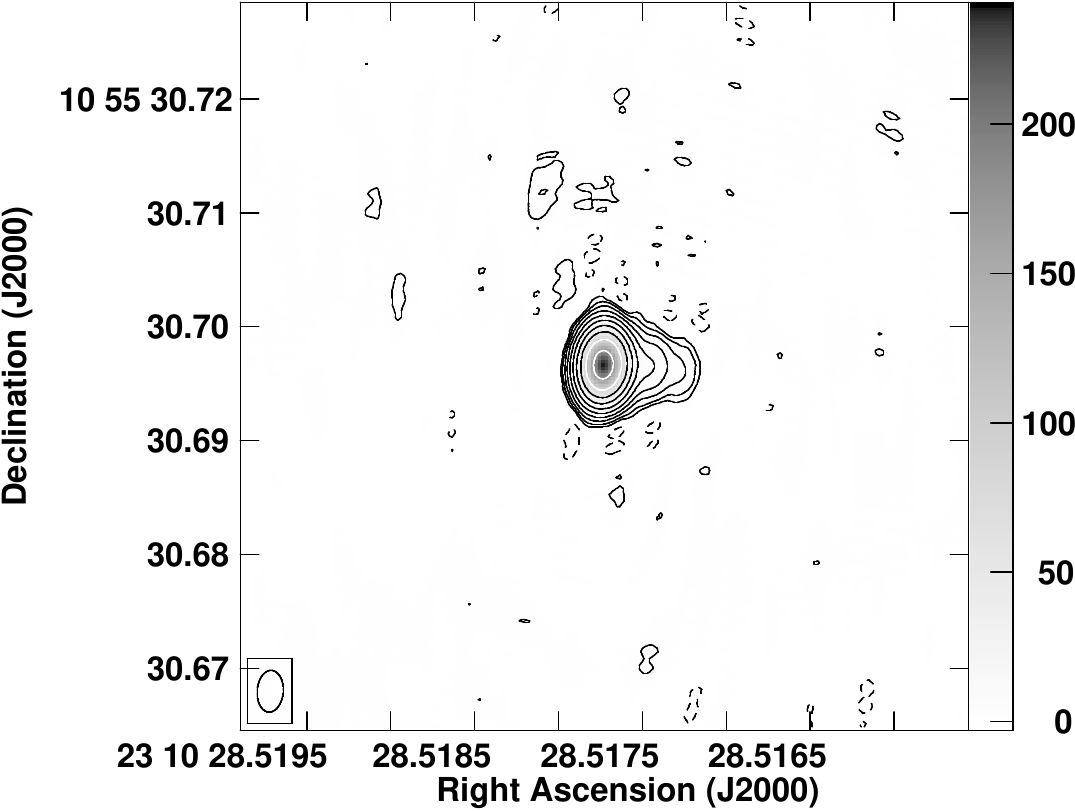}
    \includegraphics[width=0.49\linewidth]{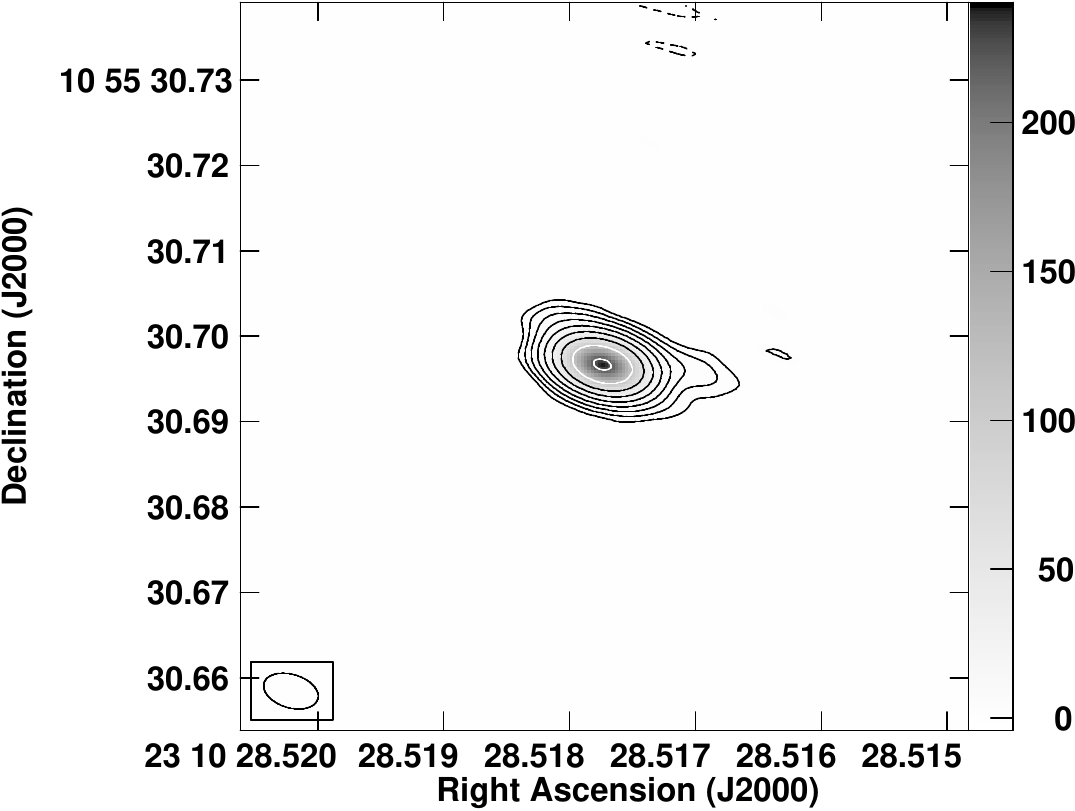}
  \end{center}

\caption{\label{evn6and18cmcal2310}6~cm (left) and 18~cm (right) EVN images of the phase calibrator J2310+1055. The contour levels are at (-3, 3, 6, 12, 24, 48, 96, 192, 384, 768, 1536) $\times$ 112.4~$\mu$Jy~beam$^{-1}$ (left) and at (-3, 3, 6, 12, 24, 48, 96, 192, 384, 768) $\times$ 288.5~$\mu$Jy~beam$^{-1}$ (right). The beams are at the bottom left corners and their sizes are 3.7 mas $\times$ 2.3 mas, P.A.= -4$^{\circ}$ (6 cm, left) and 6.6 mas $\times$ 3.9 mas, P.A.=73$^{\circ}$ (18 cm, right).}
\end{figure*}

\begin{figure*}[!htb]
  \begin{center}
    \includegraphics[width=0.49\linewidth]{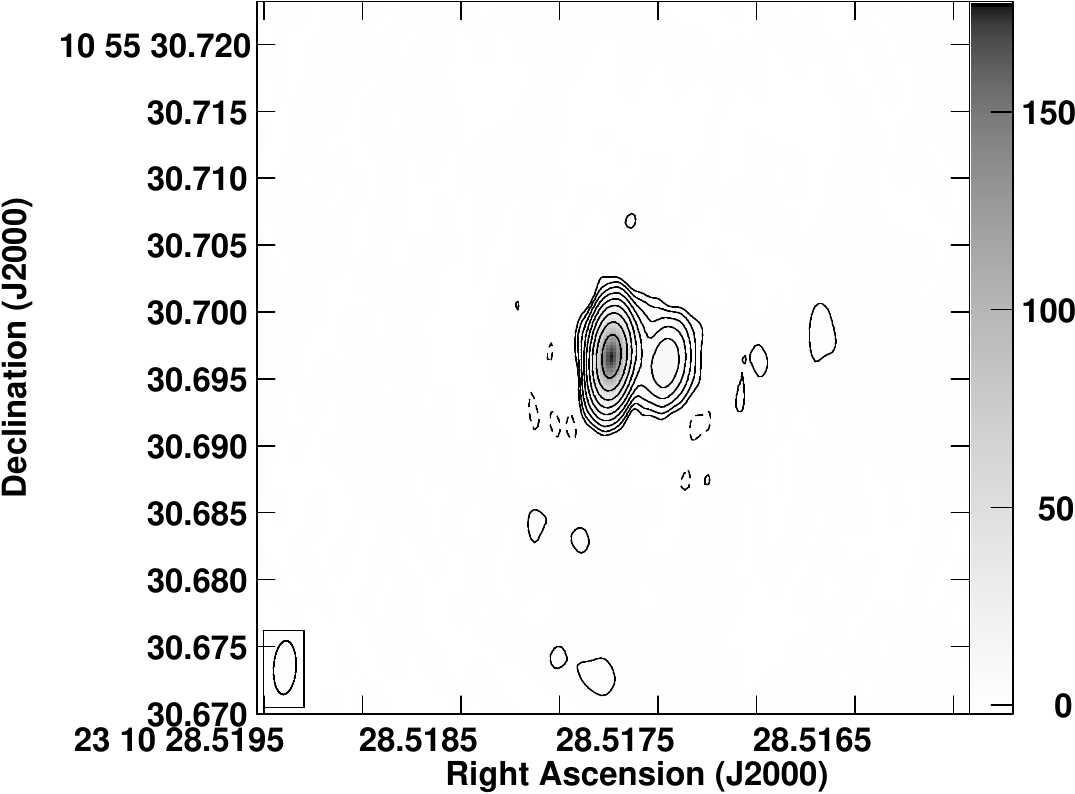}
    \includegraphics[width=0.49\linewidth]{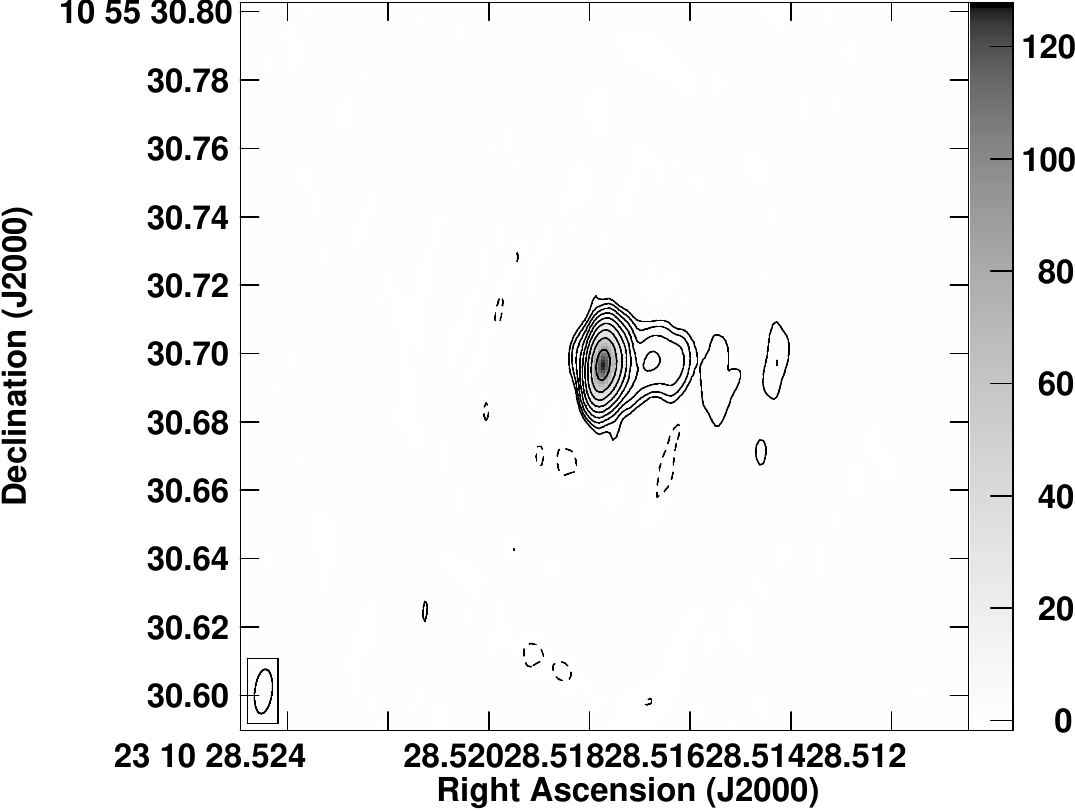}
  \end{center}

\caption{\label{vlba6and20cmcal2310}6~cm (left) and 20~cm (right) VLBA images of the phase calibrator J2310+1055. The contour levels are at (-3, 3, 6, 12, 24, 48, 96, 192, 384, 768) $\times$ 144.0~$\mu$Jy~beam$^{-1}$ (left) and at (-3, 3, 6, 12, 24, 48, 96, 192, 384, 768) $\times$ 117.2~$\mu$Jy~beam$^{-1}$ (right). The beams are at the bottom left corners and their sizes are 4.0 mas $\times$ 1.7 mas, P.A.= -4$^{\circ}$ (6 cm, left) and 13.0 mas $\times$ 5.1 mas, P.A.=-6$^{\circ}$ (20 cm, right).}
\end{figure*}

\newpage

\bibliography{ngc7479biblist}{}
\bibliographystyle{aasjournalv7}

\end{document}